\documentclass[12pt,preprint,flushrt]{aastex}
\usepackage[table]{xcolor}
\usepackage[normalem]{ulem}
\usepackage{rotating}

\begin{document}
\title{Evidence for High-Frequency QPOs with a 3:2 Frequency \\ Ratio from a 5000 Solar Mass Black Hole}
\author{Dheeraj R. Pasham$^{1,2,3}$, S. Bradley Cenko$^{1,2,3}$, Abderahmen Zoghbi$^{4,5}$, Richard 
F. Mushotzky$^{1,3}$, Jon Miller$^{4}$, Francesco Tombesi$^{1,2,3}$} \altaffiltext{1}{Astrophysics Science Division,
NASA's Goddard Space Flight Center, Greenbelt, MD 20771; email: dheerajrangareddy.pasham@nasa.gov, brad.cenko@nasa.gov} \altaffiltext{2}{Astronomy Department, University of
Maryland, College Park, MD 20742; email: ftombesi@astro.umd.edu; richard@astro.umd.edu} \altaffiltext{3}{Joint Space-Science Institute, University of Maryland, College
Park, MD 20742, USA} \altaffiltext{4}{Department of Astronomy, University of Michigan, Ann Arbor, MI 48109-1042; email: abzoghbi@umich.edu; jonmm@umich.edu} \altaffiltext{5}{Cahill Center for Astrophysics, California Institute of Technology, Pasadena, CA 91125} 

\begin{abstract}
Following the discovery of 3:2 resonance quasi-periodic 
oscillations (QPOs) in M82X-1 (Pasham et al. 2014), we 
have constructed power density spectra (PDS) of all 15 
(sufficiently long) {\it XMM-Newton} observations of the 
ultraluminous X-ray source NGC1313X-1 ($L_{X}$ $\approx$ 
2$\times$10$^{40}$ erg/sec). We detect a strong QPO at 
a frequency of 0.29$\pm$0.01 Hz in data obtained on 2012 
December 16. Subsequent searching of all the remaining 
observations for a 3:2/2:3 frequency pair revealed a 
feature at 0.46$\pm$0.02 Hz on 2003 Dec 13 (frequency 
ratio of 1.59$\pm$0.09). The global significance of the 
0.29 Hz feature considering all frequencies between 0.1 
and 4 Hz is $>$ 3.5 $\sigma$. The significance of the 
0.46$\pm$0.02 Hz QPO is $>$ 3.5$\sigma$ for a search 
at 2/3 and 3/2 of 0.29 Hz. We also detect lower 
frequency QPOs (32.9$\pm$2.6 and 79.7$\pm$1.2 mHz). All 
the QPOs are super-imposed on a continuum consisting of 
flat-topped, band-limited noise, breaking into a 
power-law at a frequency of 16$\pm$3 mHz and white 
noise at $\ga$ 0.1 Hz. NGC1313X-1's PDS is analogous to 
stellar-mass black holes' (StMBHs) PDS in the 
so-called steep power-law state, but with the respective 
frequencies (both QPOs and break frequencies) scaled down 
by a factor of $\sim$ 1000. Using the inverse 
mass-to-high-frequency QPO scaling of StMBHs, we estimate 
NGC1313X-1's black hole mass to be 5000$\pm$1300 
$M_{\odot}$, consistent with an inference from the scaling 
of the break frequency. However, the implied Eddington 
ratio, L$_{Edd}$ $>$ 0.03$\pm$0.01, is significantly lower 
compared to StMBHs in the steep power-law state (L$_{Edd}$ $\ga$ 0.2).  

\end{abstract}



\newpage
\section{Introduction}
Compact accreting X-ray sources in nearby galaxies with 
luminosities exceeding 10$^{39}$ erg s$^{-1}$ are referred 
to as ultraluminous X-ray sources (ULXs). Current evidence 
suggests that ULXs might be a mixed bag of compact objects 
including stellar-mass black holes (StMBHs: 3-25 
$M_{\odot}$) powered by super-Eddington accretion (e.g., 
King et al. 2001, Begelman 2002, Gladstone et al. 2009), 
intermediate-mass black holes (IMBHs: a few$\times$(100-1000) 
$M_{\odot}$; Kaaret et al. 2001, 2006, Matsumoto et al. 2001, Farrell et al. 
2009, Pasham et al. 2014, Mezcua et al. 2015), and neutron 
stars (Bachetti et al. 2014).

One of the biggest challenges in understanding ULXs is to 
estimate their compact object masses. Because their 
optical counterparts are faint (V-band magnitudes of 
22-24; e.g., Gladstone et al. 2013, Tao et al. 2011), Doppler 
tracking their optical counterparts to derive their mass 
functions---as done for Galactic StMBHs---has been extremely 
challenging (e.g., Roberts et al. 2011, Cseh et al. 2013). 
However, in a few ULXs, such optical measurements have 
yielded mass constraints that suggest lower-mass black holes 
($\la$ 30 $M_{\odot}$; Liu et al. 2013, Motch et al. 2014).

It has been suggested that the detection of the 3:2 frequency 
ratio high-frequency quasi-periodic oscillations (QPOs) can 
resolve the ULX mass problem (Abramowicz et al. 2004). 
StMBH high-frequency QPOs (frequency range of 100-450 Hz; 
McClintock \& Remillard 2006) that appear in a 3:2 frequency 
ratio scale inversely with the black hole mass. Moreover, it has been demonstrated that the 
power density spectra (PDS) of both the stellar-mass and the 
supermassive black holes are qualitatively similar. The PDS 
break timescale of both simply scale with the black hole mass, after accounting 
for the differences in the accretion efficiency between 
sources (e.g., McHardy et al. 2006, K{\"o}rding et al. 2007). 
One can also use this break timescale to estimate black hole 
masses. Thus, under this black hole variability unification 
paradigm, 3:2 high-frequency QPO analogs of StMBHs 
should also be detectable from IMBHs, but with centroid 
frequencies scaled down according to their respective black 
hole masses (Vaughan \& Uttley 2005). For example, a few 
1000 $M_{\odot}$ IMBH should exhibit high-frequency QPOs with 
centroid frequencies in the range of a fraction of Hz. In fact, 
such a 3:2 ratio QPOs have already been detected from the ULX 
M82 X-1. In that source, the two QPOs (3.3 and 5 Hz) allowed 
Pasham et al. (2014) to estimate its black hole mass to be 
428$\pm$105 $M_{\odot}$. Here, we report evidence for a 
second such high-frequency pair from another ULX, NGC 1313 X-1.

\section{ {\it XMM-Newton} Data}

As of the writing of this paper, 22 of the 24 {\it XMM-Newton} 
observations of NGC 1313 are public. The three brightest 
X-ray sources in this field, the two ULXs, NGC 1313 X-1 
(hereafter, X-1), X-2 and the X-ray bright supernova SN1978K 
are well separated in the XMM images (see Figure 1). Previous 
energy spectral studies of X-1 suggest it may host an 
IMBH with a mass of $\sim$ 1000 $M_{\odot}$ (e.g., Miller et 
al. 2003, 2013). Given the frequency range we are interested 
in, we used data primarily from EPIC-pn, utilizing events in 
the entire 0.3-10.0 keV band pass. Both the pn and the MOS 
detectors were operated in the so-called full-frame mode 
during all the observations. While pn's full-frame data mode 
offers a time resolution of 73.4 ms, i.e., Nyquist frequency 
of 6.82 Hz, MOS data is limited to a Nyquist frequency of 
only 0.19 Hz.

\section{Analysis}

We reduced all the observations using the standard data 
reduction procedures and removed datasets that were severely 
affected by background flaring. This preliminary screening 
left us with fifteen observations (Table 1). Source events 
were extracted from a circular region of radius 33$^{\prime\prime}$ (dashed 
circle around X-1 in Figure 1) when X-1 was clear of a CCD 
gap. When X-1 was close to or on a CCD 
gap we extracted events from a smaller region of radius 25$^{\prime\prime}$
excluding the CCD gap. Circular background regions of the 
same radius, and free of any point sources, were chosen away 
from the source's readout column and as close to the telescope 
pointing as possible. 

\subsection{Results: Timing}

In order to assess X-1's variability, we first extracted the 
source and the background light curves from each of the 
fifteen observations. Background flaring was prominent for 
only brief durations in some observations. We constructed 
good time intervals (GTIs) accounting for both the background 
flares and times when the detector was turned off. Figure 2 
contains sample background-subtracted X-ray light curves 
(black) and their respective backgrounds (red) from all the 
observations whose power spectra are described in this article.

Using the GTIs shown in Figure 2, we constructed a Leahy normalized (Poisson 
noise level of 2; Leahy 1983) PDS of X-1 from each of the 
individual observations. All the power spectra were sampled 
only up to 4 Hz, a value safely below the Nyquist frequency 
of 6.82 Hz, in order to avoid any aliasing affects. We 
started our timing analysis with the three longest 
observations (obsIDs: 0405090101, 0693850501, and 
0693851201) during which X-1 was positioned on-axis, giving 
the best sensitivity for detecting QPOs. 

\subsubsection{ObsID 0405090101}

The top panel of Figure 3 shows the combined EPIC (pn+MOS) PDS 
from obsID 0405090101---sampled in the frequency range from 
0.0022 to 4 Hz. We first extracted a pn-only PDS and found a 
QPO-like feature at 80 mHz, a frequency well below the MOS 
detectors' Nyquist frequency. Therefore, in order to improve 
the signal, we used pn+MOS data in this single instance. It 
is evident that the overall shape of the power spectrum is 
flat-topped at the lowest frequencies, breaking into a 
power-law, and white noise at the highest frequencies. Two 
broad, QPO features at centroid frequencies of roughly 30 
and 80 mHz are also apparent. We modeled the continuum with 
a constant plus a bending power law model$\footnote{\[Power = C + \frac{N 
\times \nu^{-\alpha} }{\left[1+ \left(\frac{\nu}{\nu_{bend}} 
\right)^{\beta-\alpha}\right]} \] where C, N, $\nu_{bend}$ are 
the Poisson noise level, the normalization of the bending 
power-law and the bending frequency, respectively while 
$\alpha$ and $\beta$ are the power-law indicies below and above 
the bending frequency, respectively.}$ similar to other ULXs 
(e.g., Pasham \& Strohmayer 2012, 2013), and StMBHs in the steep 
power-law state (McClintock \& Remillard 2006). This gave a best-fit 
$\chi^2$ of 419 for 252 degrees of freedom (dof). We then 
added a Lorentzian component to model the QPO at 80 mHz. 
This improved the $\chi^2$ by 89, i.e., $\chi^2$ of 330 for 
249 dof. Using the F-test, this corresponds to a significance 
of 8$\times$10$^{-13}$ or $>$ 7$\sigma$. The best-fit QPO 
has a centroid frequency, width, normalization, and a 
root-mean-squared (RMS) amplitude of 79.7$\pm$1.2 mHz, 
14.5$\pm$3.4 mHz, 0.57$\pm$0.09, and 10.8$\pm$2.4\%, 
respectively. Adding a second QPO improved the $\chi^2$ by 40 
(290 for 246 dof), which corresponds to an F-test probability 
of 6$\times$10$^{-7}$ ($>$ 4.9$\sigma$). The best-fit centroid 
frequency, width, normalization and RMS amplitude of the 
second QPO were 32.9$\pm$2.6 mHz, 18.2$\pm$7.8, 0.34$\pm$0.08, 
and 9.4$\pm$3.9\%, respectively. 

However, Protassov et al. (2002) pointed out the problems with 
applying the F-test in additive models. Therefore, in order 
to estimate the QPO significances independent 
of the F-test, we employed a rigorous Monte Carlo approach 
as follows.

1) We estimated the baseline bending power-law plus a 
constant model parameters along with their uncertainties, 
and then randomly sampled $N$ = 1.8$\times$10$^{6}$ model 
parameter sets from within the best-fit parameter error 
bars (assuming normal distribution). 

2) For each of these parameter sets, we simulated a light 
curve of the same length as the observed one following the 
algorithm described in Timmer \& Koenig (1995), and then 
extracted a PDS from this simulated light curve.

3) After binning these PDS in the same way as the original 
PDS, we modeled them with a bending power law. We then added 
a QPO to this model, with the QPO frequency constrained to 
lie between 0.01 and 1 Hz. The maximum  improvement in 
$\chi^2$ ($\Delta$$\chi^{2}_{max}$) was recorded from each 
simulated PDS.

4) The significance of the 80 mHz QPO was estimated as 1 - 
$N_{(\Delta\chi^{2}_{max} > \Delta\chi^2_{obs})}$/$N$, where 
$N_{(\Delta\chi^{2}_{max} > \Delta\chi^2_{obs})}$ is the 
number of simulated $\Delta\chi^{2}_{max}$ values greater 
than the observed $\Delta\chi^2_{obs}$. For estimating the 
significance of the 30 mHz QPO, we assumed the baseline 
model to be the best-fit continuum plus 80 mHz QPO of the 
observed PDS, and recorded the maximum $\Delta\chi^2$ 
values by adding an additional QPO to this base model.

This methodology is similar to estimating the significances 
of spectral lines in energy spectra (e.g., Tombesi et al. 
2010, Zoghbi et al. 2015). For the 80 mHz feature we ran 
1.8$\times$10$^{6}$ simulations. The maximum $\Delta\chi^2$ 
was 25, which is much lower than the observed value of 89. 
Thus, we conclude that the 80 mHz feature is significant 
at least at the 5$\sigma$ level. We ran 200,000 simulations 
to test the significance of the 30 mHz feature and found 
that one run exceeded the observed $\Delta\chi^2$ of 40. 
Thus, we conclude its significance is 1 - (1/200,000) or 
$\approx$ 4.4$\sigma$. 

\subsubsection{ObsID 0693850501}

The PDS from observation 0693850501 exhibited a strong 
feature at roughly 0.3 Hz (bottom-left panel of Figure 3). 
The continuum looks flat because we only sampled up to 
roughly 0.02 Hz. A flat-topped, followed by a power-law like 
continuum can be seen when we sample up to 0.001 Hz using 
longer light curve segments. Modeling the continuum with 
a constant yielded a best-fit $\chi^2$ 
of 186 for 127 dof. Adding a Lorentzian component improved 
the $\chi^2$ by 56 with an addition of three parameters 
($\chi^2$ of 130 for 124 dof). This corresponds to an F-test 
probability of 1.1$\times$10$^{-9}$ or $>$ 6$\sigma$. Using 
the Monte Carlo approach to test the significance, with 
1.8$\times$10$^{6}$ simulations, we find a lower limit of 
5$\sigma$. For the Monte Carlo simulations, we modeled each 
of the simulated PDS with a constant, and a constant plus 
a Lorentzian model, and recorded the maximum improvement in 
the $\Delta\chi^2$. Out of the 1.8$\times$10$^{6}$ simulations,
the maximum $\Delta\chi^2$ improvement was 27, a value much 
less than the observed $\Delta\chi^2$ improvement of 56. The 
best-fit centroid frequency, width, normalization and the RMS 
of the QPO were 0.29$\pm$0.01 Hz, 0.13$\pm$0.04 Hz, 
0.14$\pm$0.03, and 19.0$\pm$5.0\%, respectively.

We also estimated the significance using another independent 
method. First we rescaled the PDS (rescaling factor of 1.01, 
and the rescaled power at 0.29 Hz--in the highest bin--is 2.173) 
so that the local mean around 0.5 Hz is equal to 2, the value expected from a 
purely Poisson (white noise) process. We then computed the 
probability with 3$\sigma$ and the 4$\sigma$ confidence, of
obtaining a power that is at or higher than some threshold 
P$_{*}$, Probability(P $>$ P$_{*}$) = N$_{trials}$$\times$Q(P$_{*}$$\times$898$\times$4$|$2$\times$898$\times$4), 
where Q(P$_{*}$$\times$898$\times$4$|$2$\times$898$\times$4) 
is the probability of obtaining a $\chi^2$ value of 
P$_{*}$$\times$898$\times$4 or higher from a $\chi^2$ 
distribution with 2$\times$898$\times$4 dof. We used this 
$\chi^2$ distribution because we averaged in frequency by a 
factor of 4 and averaged 898 individual power spectra each 
derived from 128 s light curve segments. N$_{trials}$ account for
the total number of trials (frequency bins within 0.1-4 Hz).
The confidence contours are marked by horizontal dotted 
lines in the bottom-left panel of Figure 3. Clearly, the 
highest bin in the 0.3 Hz QPO is significant at greater 
than the 4$\sigma$ level.

After establishing the lower-frequency continuum and QPOs, 
and the 0.3 Hz QPO, we searched for a signal at 2/3 and 3/2 
of 0.3 Hz separately from all the GTIs longer than 7 ks.
The PDS from the 4$^{th}$ GTI (exposure $\approx$ 22 ks) 
of this observation (top panel of Figure 4) showed evidence 
for excess power at 0.44$\pm$0.06 Hz, a value consistent 
with 3/2 of 0.3 Hz. This feature is significant at the 
7.6$\times$10$^{-4}$ level (the significance is 1$\times$10$^{-2}$
if all frequencies were searched). We estimated the significance 
of this feature as follows. First, we estimated the 
probability of detecting a false peak with a power value 
of 2.13. This is the probability of getting a $\chi^2$ 
value of 2.13$\times$174$\times$16 or higher from a $\chi^2$ 
distribution with 2$\times$174$\times$16 dof. This value 
is 3.8$\times$10$^{-4}$. However, after securing the 0.3 
Hz feature, we searched in two bins, one at 2/3 and 
the other at 3/2 (bin width of 0.125 Hz). Considering the 
two trials, the significance in the 4$^{th}$ GTI is 
2$\times$3.8$\times$10$^{-4}$. This significance level 
does not take into account the number of GTIs searched 
yet. We estimate its global significance--considering 
all GTIs--in section 3.1.5. 

\subsubsection{ObsID 0693851201}

A feature at a frequency of 0.30$\pm$0.02 Hz (middle panel of 
Figure 4) as observed in the PDS of observation 0693850501 
(taken roughly a week before this observation) was again 
present, albeit at a lower significance of $\ga$ 3.3$\sigma$. 
Similar to the bottom-left panel of Figure 3, the dotted 
horizontal lines represent the 3 and the 4$\sigma$ confidence 
contours considering all frequency bins (trials) between 0.1
 and 4 Hz.

\subsubsection{Other Observations}

We then constructed PDS from each of the shorter observations. 
Observation 0150280401 showed evidence for a QPO with a 
centroid frequency of 0.46$\pm$0.02 Hz (RMS amplitude of 
12$\pm$2\%), a value consistent with 3/2 of 0.3 Hz. The 
significance of this feature--again considering a search 
around 2/3 and 3/2 of 0.3 Hz--is 2.8$\times$10$^{-4}$
estimated as follows. First, we estimated the probability 
of getting a power value of 2.451, i.e., the probability of  
getting a $\chi^2$ value of 2.451$\times$81$\times$4 or higher 
from a $\chi^2$ distribution with 2$\times$81$\times$4 dof. 
This value is 7$\times$10$^{-5}$. However, considering that we  
searched in two bins (width per bin of 0.03125 Hz) around 3/2 
and 2/3 of 0.3 Hz, this translates to 4$\times$7$\times$10$^{-5}$ 
($>$ 3.5$\sigma$). The significance of the feature considering 
a full frequency search is 6$\times$10$^{-3}$.

Observation 0205230601 showed weak evidence for QPO-like feature
at the $>$ 2$\sigma$ level (Bottom panel of Figure 4). The PDS 
from the remaining observations were essentially featureless with evidence for red-noise in a 
handful of them. To ensure that these QPO features are 
not associated with the background, we extracted all the 
background PDS from these fifteen observations. 
They are all consistent with being featureless (flat, 
white noise) within the error bars.

\subsubsection{Global probability of the 0.45 and the 0.3 Hz QPOs}
In order to estimate the global significance of the 0.45 Hz
QPO we considered all the 24 GTIs--longer than 7 ks--where we 
searched for QPOs (see Table 1). Figures 3 \& 4 show all the 
statistically significant QPOs detected during this search.
The QPO's global probability can be calculated straightforwardly using 
the binomial distribution formula which gives the 
probability of happening of a certain event $m$ times in 
$n$ trials as,  

\[P(m; p,n) = \frac{n!}{m!(n-m)!}p^m(1-p)^{n-m}\]

where $n$ is the total number of GTIs searched, and 
$m$ is the number of GTIs where the signal was detected 
at a probability of $p$.

Using the above formula, the global probability of 
detecting the 0.45 Hz feature at $>$ 7.6$\times$10$^{-4}$ 
significance in two ($m$) out of the twenty four ($n$) GTIs is 
1.6$\times$10$^{-4}$ ($>$ 3.5$\sigma$). 

Similarly, we estimate the global probability of the 0.3 Hz
QPO, using $n$ = 24, $m$ = 1, and $p$ = 5.6$\times$10$^{-7}$ 
(5$\sigma$), to be $>$ 3.5$\sigma$. {\it Note that this is very
conservative lower limit as we have not included the case 
where the 0.3 Hz feature was detected at $>$ 3.3$\sigma$}.

\section{Discussion}
NGC 1313 X-1's PDS has all the features of a typical 
StMBH in the steep power-law state, but with characteristics 
time scales $\sim$1000$\times$ longer. Typical StMBHs have  
three components: (1) A continuum often flat-topped at the 
lowest frequencies, with a power-law like decline beyond a 
certain break frequency, followed by white noise at the 
highest frequencies; (2) Low-frequency QPOs (frequencies of 
a few Hz); and, finally (3) High-frequency QPOs (frequency 
range of 100-450 Hz) exhibited by some systems. Three StMBHs 
with known masses show high-frequency QPOs in harmonic pairs 
with centroid frequencies in a ratio consistent with 3:2 
(e.g., Miller et al. 2001, Strohmayer et al. 2001a, 2001b, 
Remillard et al. 2002). In these systems, the two QPOs are 
often not simultaneous (e.g., see Table 1 of Remillard 2002 
\& Strohmayer 2001b). Furthermore, unlike the low-frequency 
QPOs, these are stable in frequency with changes in source 
luminosity (McClintock \& Remillard 2006). Also, the 
timescales associated with them ($\sim$0.01 s) are comparable 
to the Keplerian orbital periods of a test particle close to 
the innermost stable circular orbit. The commonalities of the 
high-frequency QPOs in these three StMBH systems suggests a 
common physical origin. Under the assumption that these 
originate from a radius fixed in gravitational units ($GM$/$c^{2}$, 
where $G$, $M$, and $c$ are the Gravitational constant, the 
black hole mass and the speed of light, respectively), their 
frequency should simply scale inversely with the black hole 
mass. Indeed, the three StMBHs with high-frequency QPOs do 
agree with this inverse scaling law (see Figure 4.17 of 
McClintock \& Remillard 2006, Zhou et al. 2015).

We suggest that the observed lower 30 and 80 mHz, and the 
higher 0.3 and 0.45 Hz 3:2 ratio QPOs are the analogs of 
the low and the high-frequency QPOs of StMBHs, and that 
X-1 may be in an X-ray accretion state similar to the steep 
power-law state of StMBHs. This result agrees with prior 
work by Feng \& Kaaret (2006), who studied X-1's X-ray 
(0.3-10 keV) energy spectral variability using {\it 
XMM-Newton} data taken between 2000 and 2005. They concluded 
that the source resides in either the steep power-law 
state--at high luminosities--or in the low/hard state at 
lower luminosities, but never enters the thermal dominant 
state (see Bachetti et al. 2013 for alternate arguments).

Using the dynamical masses and the high-frequency QPOs 
of StMBHs$\footnote{XTE J1550-564, GRO J1655-50 and 
GRS 1915+105 have high-frequency QPOs at 184 \& 276 Hz,
300 \& 450 Hz, and 113 \& 168 Hz, respectively. Their
black hole masses are 9.1$\pm$0.61 $M_{\odot}$ (Orosz et 
al. 2011), 5.4$\pm$0.3 $M_{\odot}$ (Beer \& Podsiadlowski 
2002), and 10.1$\pm$0.6 $M_{\odot}$ (Steeghs et al. 2013),
respectively.}$ XTE J1550-564, GRO J1655-50, and GRS 1915+105,
we estimate---based on the inverse mass scaling---that X-1's
black hole mass is 5000$\pm$1300 $M_{\odot}$. We also 
measured the mass using the break frequency--black hole 
mass-accretion rate scaling as derived by McHardy et al. 
(2006). Using a break frequency of 16$\pm$3 mHz from obsID 
0405090101, and assuming a lower limit on the bolometric 
luminosity of 2$\times$10$^{40}$ erg s$^{-1}$ implies a
black hole mass of $>$ 7600$\pm$5600 $M_{\odot}$$\footnote{The 
high uncertainty from the break frequency scaling 
is due to large error bars on the coefficients of the scaling 
law; See Figure 1 of McHardy et al. (2006).}$. This value is 
consistent with the measurement from the 3:2 QPO pair. 

Yang et al. (2011) estimated X-1's bolometric correction 
factor (ratio of the X-ray--to--Optical flux) to be similar 
to that of typical StMBHs, implying that 
a significant fraction of the emission is in the X-rays. X-1's 
average X-ray (0.3-10.0 keV) luminosity of 2$\times$10$^{40}$ 
erg s$^{-1}$ (e.g., Feng \& Kaaret 2006) implies an Eddington 
ratio of $>$ 0.03$\pm$0.01. This value is significantly low 
compared to the typical Eddington ratios of StMBHs in the steep 
power-law state ($>$ 0.2$L_{Edd}$; McClintock \& Remillard et 
al. 2006). Assuming the 0.2 value reported by McClintock \& 
Remillard (2006), X-1's low Eddington ratio is inconsistent 
with the interpretation of it being in the steep power-law state.

Bachetti et al. (2013) carried out timing analysis of  
two of the data sets described here (0693850501 and 
0693851201). However, they did not report any evidence 
for QPOs at 0.3 Hz. We suspect 
the reason for this discrepancy is because they 
extracted their PDS using both the pn and the MOS data. 
As described earlier, MOS is limited to a frequency of 
$\approx$ 0.19 Hz. In addition to any signal beyond 
0.19 Hz being unreliable, even at frequencies lower than, 
but close to 0.19 Hz, signal suppression can be severe 
(van der Klis 1989). To test this, we extracted a PDS 
using a combined pn and MOS event list from the 
observations 0693850501 and 0693851201. While we still 
see evidence for a feature at 0.3 Hz and and at $\sim$ 
90 mHz, they were statistically less significant than 
in the analysis presented above.

The first ULX 3:2 pair from M82 X-1 was detected using 
the Rossi X-ray Timing Explorer ({\it RXTE}), while the 
detection reported here is from {\it XMM-Newton}. 
Currently, {\it XMM-Newton} is the only X-ray observatory 
that can provide both large enough effective area and the 
required time resolution to detect these oscillations 
from ULXs. Therefore, deeper X-ray observations of other 
variable ULXs, {\it viz.}, NGC 5408 X-1 (Middleton 
et al. 2011), NGC 6946 X-1 (Rao et al. 2010) are strongly encouraged. Also, the 
reported 3:2 pair from X-1 boosts confidence 
in the prospects of detecting high-frequency QPOs from
relatively isolated ULXs with 
Neutron star Interior Composition ExploreR ({\it NICER})---with 
an anticipated effective area of 1.5 times greater
than EPIC-pn.

{\bf Acknowledgements:} We thank the referee for his/her valuable 
comments/suggestions that improved the paper.

\vfill\eject

\begin{table}
    \caption{{ Summary of the {\it XMM-Newton} observations of NGC 1313 X-1}}\label{Table1} 
{\small
\begin{center}
   \begin{tabular}[t]{lccccccc}
	\hline\hline \\
    ObsID	& Date (UTC)\tablenotemark{1}  	&	Observation			 	& Count rate\tablenotemark{3}	& Effective 				& Number of  \\
		& 				&	Time (ks)\tablenotemark{2}              & (counts s$^{-1}$)             	& Exposure\tablenotemark{4}(ks) & GTIs $>$ 7 ks\\
	\\
    \hline \\
 0106860101 	&  2000-10-17 			&	42.4 					& $0.73 \pm 0.005$  			& 31.6  			& 	1		\\
\\
 0150280301  	&  2003-12-21 			&	16.2					& $1.02 \pm 0.01$  			& 9.7 				&	1	 \\
\\
 0150280401  	&  2003-12-23			&	20.9					& $0.91 \pm 0.008$  			& 10.5 				&	1		 \\
\\
 0150280501  	&  2003-12-25			&	21.4					& $0.70 \pm 0.007$  			& 9.8 				&	1		 \\
\\
 0150280601  	&  2004-01-08 			&	53.2					& $0.79 \pm 0.007$  			& 12.4 				&	1		 \\
\\
0150281101  	&  2004-01-16 			&	8.9					& $0.87 \pm 0.01$  			& 7.0 				&	1		 \\
\\
 0205230301	&  2004-06-05			&	11.9					& $1.27 \pm 0.01$  			& 10.0 				&	1		 \\
\\
 0205230401  	&  2004-08-23			&	18.0					& $0.63 \pm 0.006$  			& 14.9 				&	1		 \\
\\
0205230501  	&  2004-11-23			&	16.0					& $0.26 \pm 0.004$  			& 14.0 				&	1		 \\
\\
 0205230601  	&  2005-02-07			&	14.3					& $0.57 \pm 0.007$  			& 12.4 				&	1		 \\
\\	
 0301860101  	&  2006-03-06			&	21.8					& $1.11 \pm 0.008$  			& 19.9 				&	1		 \\
\\
 0405090101  	&  2006-10-15			&	123.1					& $0.70 \pm 0.002$  			& 121.1 			&	3		 \\
\\
 0693850501 	&  2012-12-16			&	125.2					& $0.83 \pm 0.003$  			& 123.0 			&	4		 \\
\\
 0693851201  	&  2012-12-22			&	125.2					& $0.85 \pm 0.003$  			& 123.0 			&	4		 \\
\\
 0722650101  	&  2013-06-08			&	30.7 					& $0.71 \pm 0.005$  			& 28.8 				&	2		 \\
\\
    \hline\hline
    \end{tabular}
\end{center}
}
\tablenotemark{1}{Coordinated Universal Time.}
\tablenotemark{2}{Total observation time in ks.}
\tablenotemark{3}{Average EPIC-pn 0.3-10 keV count rate of NGC 1313 X-1. Note, however, that X-1 was not always on-axis.}
\tablenotemark{4}{After accounting for flaring background and instrumental good time intervals.}\\
\end{table}
\vfill\eject


\begin{figure}[ht]

\begin{center}
\includegraphics[width=4.75in, height=4.35in, angle=0]{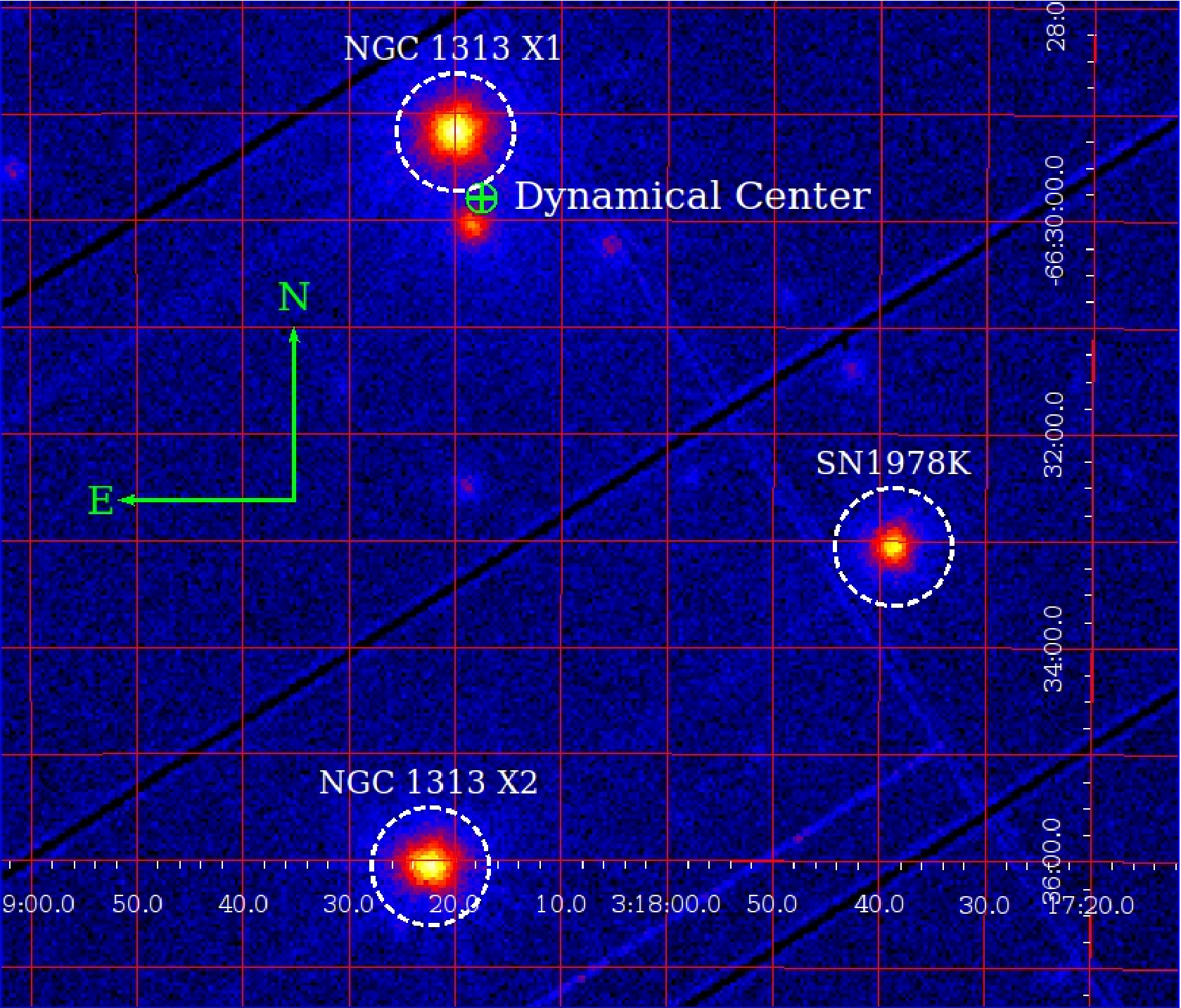}
\end{center}
{\textbf{Figure 1:} {\it XMM-Newton}/EPIC-pn X-ray (0.3-10 keV) image of NGC 
1313, from observation 0693850501, produced with DS9. The dashed circles 
have a radius of 33$^{\prime\prime}$ and indicate the size of our typical source extraction region. The 
dynamical center of the host galaxy, shown as an encircled cross, is based 
on H I maps by Ryder et al. (1995). The north and the east arrows are each 
100$^{\prime\prime}$ long.}
\label{fig:figure1}
\end{figure}


\begin{figure}[ht]

\begin{center}
\includegraphics[width=6.in, height=7.5in, angle=0]{Figure2.ps}
\end{center}
{\textbf{Figure 2:} Background-subtracted EPIC-pn X-ray (0.3-10 keV) light curves of NGC 1313 X-1 
(black), and their backgrounds (red) during five different {\it XMM-Newton} 
observations. The PDS from these data sets are shown in Figures 3 \& 4. 
For each panel, their observation ID and time zero in seconds since 1998.0 
TT are indicated at the top. Also shown are the are the good time intervals 
(GTIs) with duration longer than 500 s. A dashed vertical green line and a 
solid vertical blue line mark the beginning and end of a GTI, respectively. 
}
\label{fig:figure1}
\end{figure}


\begin{figure}[ht]

\begin{center}
\hspace{-2cm}
\includegraphics[width=6.in, height=4.75in, angle=0]{Figure3.ps}
\end{center}
\vspace{-.35cm} 
{\textbf{Figure 3:} {\it XMM-Newton}/EPIC (0.3-10.0 keV) power density spectra 
of NGC 1313 X-1 from three different observations. {\bf Top Left Panel:} pn+MOS PDS 
from observation 0405090101 (frequency resolution of 3.9$\times$10$^{-3}$ Hz). 
The continuum was modeled with a constant plus a bending power-law, while the 
QPOs at 30 mHz (Q value: $\nu$/$\Delta\nu$ $\approx$ 2) and 80 mHZ (Q value 
$\approx$ 5.5) were modeled with Lorentzians. The solid blue line is the 
best-fit model. The PDS was obtained by averaging 235 PDS constructed from 512 
s light curve segments binned at 1/8 s. {\bf Top Right Panel:} pn only PDS from
observation 0405090101 (frequency resolution of 7.8$\times$10$^{-3}$ Hz). The 
3 and the 4$\sigma$ contours for $\ga$ 0.1 Hz (considering all frequency bins 
between 0.1 and 4 Hz), and the best-fit bending power-law
model with a QPO (blue) are also shown. {\bf Bottom Left Panel:} EPIC-pn PDS 
(frequency resolution of 0.03125 Hz) from observation 0693850501 showing a strong 
QPO feature at 0.29$\pm$0.01 Hz (Q value = 2.2). Also shown are the 3 and the 4$\sigma$ significance contours considering
all the frequency bins (trials) within 0.1 and 4 Hz of this observation. While the F-test suggests 
a significance of $>$ 6$\sigma$, full Monte Carlo simulations imply a lower 
limit of 5$\sigma$. {\bf Bottom Right Panel:} EPIC-pn PDS from observation 
0150280401 showing a power spectral feature at 0.46$\pm$0.02 Hz (Q $>$ 14), 
consistent with 3/2 of 0.29 Hz. This PDS was constructed from 128 s segments 
binned at 1/8 s. The 3 and the 4$\sigma$ contours take into account 4 bins (2
at 3/2 and 2/3 of 0.3 Hz and each 0.03125 Hz wide). In all the three cases, 
the Poisson noise level is marked by a dashed red line and the 1$\sigma$ error 
bars of the highest bins of the QPOs are indicated in gray.}
\label{fig:figure1}
\end{figure}


\begin{figure}[ht]

\begin{center}
\hspace{-2cm}
\includegraphics[width=2.7in, height=6.75in, angle=0]{Figure4.ps}
\end{center}
\vspace{-.35cm} 
{\textbf{Figure 4:} EPIC-pn (0.3-10.0 keV) power 
spectra of NGC 1313 X-1 from observations 0693850501 (top: using only the 
last GTI of $\approx$ 22 ks), 0693851201 (middle), and 0205230601 (bottom). 
These show epochs--in addition to those shown in Figure 3--with excess 
power around 0.3 and 0.45 Hz. The frequency resolution in the top, middle 
and the bottom PDS is 0.125, 0.03125 and 0.03125 Hz, respectively. The top 
PDS was constructed from 128 s light curve segments while the middle and 
the bottom were constructed from 256 s light curve segments binned at 1/8 
s. The number of PDS averaged in the top, middle and the bottom panels were 
174, 478, and 48, respectively. The confidence contours (estimated for a 
given observation and are not global; see 3.1.5 for global significances) 
in the top and the bottom panels take into account two, and four frequency 
bins, respectively (see text), while the contours in the middle panel take 
all the frequency bins between 0.1 and 4 Hz into account.
}
\label{fig:figure1}
\end{figure}



\begin{thebibliography}

\bibitem[Abramowicz et al.(2004)]{2004ApJ...609L..63A} Abramowicz, M.~A., Klu{\'z}niak, W., McClintock, J.~E., \& Remillard, R.~A.\ 2004, \apjl, 609, L63 

\bibitem[Abramowicz \& Klu{\'z}niak(2004)]{2004AIPC..714...21A} Abramowicz, M.~A., \& Klu{\'z}niak, W.\ 2004, X-ray Timing 2003: Rossi and Beyond, 714, 21 

\bibitem[Bachetti et al.(2013)]{2013ApJ...778..163B} Bachetti, M., Rana, V., Walton, D.~J., et al.\ 2013, \apj, 778, 163 

\bibitem[Bachetti et al.(2014)]{2014Natur.514..202B} Bachetti, M., Harrison, F.~A., Walton, D.~J., et al.\ 2014, \nat, 514, 202 

\bibitem[Beer \& Podsiadlowski(2002)]{2002MNRAS.331..351B} Beer, M.~E., \& Podsiadlowski, P.\ 2002, \mnras, 331, 351 

\bibitem[Begelman(2002)]{2002ApJ...568L..97B} Begelman, M.~C.\ 2002, \apjl, 568, L97 

\bibitem[Belloni et al.(2012)]{2012MNRAS.426.1701B} Belloni, T.~M., Sanna, A., \& M{\'e}ndez, M.\ 2012, \mnras, 426, 1701 

\bibitem[Cseh et al.(2013)]{2013MNRAS.435.2896C} Cseh, D., Gris{\'e}, F., Kaaret, P., et al.\ 2013, \mnras, 435, 2896 

\bibitem[Dheeraj \& Strohmayer(2012)]{2012ApJ...753..139D} Dheeraj, P.~R., \& Strohmayer, T.~E.\ 2012, \apj, 753, 139 

\bibitem[Farrell et al.(2009)]{2009Natur.460...73F} Farrell, S.~A., Webb, N.~A., Barret, D., Godet, O., \& Rodrigues, J.~M.\ 2009, \nat, 460, 73 

\bibitem[Feng \& Kaaret(2006)]{2006ApJ...650L..75F} Feng, H., \& Kaaret, P.\ 2006, \apjl, 650, L75 

\bibitem[Gladstone et al.(2009)]{2009MNRAS.397.1836G} Gladstone, J.~C., Roberts, T.~P., \& Done, C.\ 2009, \mnras, 397, 1836 

\bibitem[Gladstone et al.(2013)]{2013ApJS..206...14G} Gladstone, J.~C., Copperwheat, C., Heinke, C.~O., et al.\ 2013, \apjs, 206, 14 

\bibitem[Kaaret et al.(2001)]{2001MNRAS.321L..29K} Kaaret, P., Prestwich, A.~H., Zezas, A., et al.\ 2001, \mnras, 321, L29 

\bibitem[Kaaret et al.(2006)]{2006Sci...311..491K} Kaaret, P., Simet, M.~G., \& Lang, C.~C.\ 2006, Science, 311, 491 

\bibitem[King et al.(2001)]{2001ApJ...552L.109K} King, A.~R., Davies, M.~B., Ward, M.~J., Fabbiano, G., \& Elvis, M.\ 2001, \apjl, 552, L109 

\bibitem[K{\"o}rding et al.(2007)]{2007MNRAS.380..301K} K{\"o}rding, E.~G., Migliari, S., Fender, R., Belloni, T., Knigge, C., \& McHardy, I.\ 2007, \mnras, 380, 301 

\bibitem[Leahy et al.(1983)]{1983ApJ...266..160L} Leahy, D.~A., Darbro, W., Elsner, R.~F., et al.\ 1983, \apj, 266, 160 

\bibitem[Liu et al.(2013)]{2013Natur.503..500L} Liu, J.-F., Bregman, J.~N., Bai, Y., Justham, S., \& Crowther, P.\ 2013, \nat, 503, 500 

\bibitem[Matsumoto et al.(2001)]{2001ApJ...547L..25M} Matsumoto, H., Tsuru, T.~G., Koyama, K., et al.\ 2001, \apjl, 547, L25 

\bibitem[McClintock \& Remillard(2006)]{2006csxs.book..157M} McClintock, J.~E., \& Remillard, R.~A.\ 2006, Compact stellar X-ray sources, 157 

\bibitem[McHardy et al.(2006)]{2006Natur.444..730M} McHardy, I.~M., Koerding, E., Knigge, C., Uttley, P., \& Fender, R.~P.\ 2006, \nat, 444, 730 

\bibitem[Mezcua et al.(2015)]{2015MNRAS.448.1893M} Mezcua, M., Roberts, T.~P., Lobanov, A.~P., \& Sutton, A.~D.\ 2015, \mnras, 448, 1893 


\bibitem[Miller et al.(2001)]{2001ApJ...563..928M} Miller, J.~M., Wijnands, R., Homan, J., et al.\ 2001, \apj, 563, 928 

\bibitem[Miller et al.(2003)]{2003ApJ...585L..37M} Miller, J.~M., Fabbiano, G., Miller, M.~C., \& Fabian, A.~C.\ 2003, \apjl, 585, L37 

\bibitem[Miller et al.(2004)]{2004ApJ...614L.117M} Miller, J.~M., Fabian, A.~C., \& Miller, M.~C.\ 2004, \apjl, 614, L117 

\bibitem[Miller et al.(2013)]{2013ApJ...776L..36M} Miller, J.~M., Walton, D.~J., King, A.~L., et al.\ 2013, \apjl, 776, L36 

\bibitem[Motch et al.(2014)]{2014Natur.514..198M} Motch, C., Pakull, M.~W., Soria, R., Gris{\'e}, F., \& Pietrzy{\'n}ski, G.\ 2014, \nat, 514, 198 


\bibitem[Orosz et al.(2011)]{2011ApJ...730...75O} Orosz, J.~A., Steiner, J.~F., McClintock, J.~E., et al.\ 2011, \apj, 730, 75 

\bibitem[Pasham \& Strohmayer(2013)]{2013ApJ...771..101P} Pasham, D.~R., \& Strohmayer, T.~E.\ 2013, \apj, 771, 101 

\bibitem[Pasham et al.(2014)]{2014Natur.513...74P} Pasham, D.~R., Strohmayer, T.~E., \& Mushotzky, R.~F.\ 2014, \nat, 513, 74 

\bibitem[Protassov et al.(2002)]{2002ApJ...571..545P} Protassov, R., van Dyk, D.~A., Connors, A., Kashyap, V.~L., \& Siemiginowska, A.\ 2002, \apj, 571, 545 


\bibitem[Remillard et al.(2002)]{2002ApJ...580.1030R} Remillard, R.~A., Muno, M.~P., McClintock, J.~E., \& Orosz, J.~A.\ 2002a, \apj, 580, 1030 

\bibitem[Remillard et al.(2006)]{2006ApJ...637.1002R} Remillard, R.~A., McClintock, J.~E., Orosz, J.~A., \& Levine, A.~M.\ 2006, \apj, 637, 1002 

\bibitem[Roberts et al.(2011)]{2011AN....332..398R} Roberts, T.~P., Gladstone, J.~C., Goulding, A.~D., et al.\ 2011, Astronomische Nachrichten, 332, 398 


\bibitem[Ryder et al.(1995)]{1995AJ....109.1592R} Ryder, S.~D., Staveley-Smith, L., Malin, D., \& Walsh, W.\ 1995, \aj, 109, 1592 

\bibitem[Steeghs et al.(2013)]{2013ApJ...768..185S} Steeghs, D., McClintock, J.~E., Parsons, S.~G., et al.\ 2013, \apj, 768, 185 


\bibitem[Strohmayer(2001a)]{2001ApJ...552L..49S} Strohmayer, T.~E.\ 2001a, \apjl, 552, L49 

\bibitem[Strohmayer(2001b)]{2001ApJ...554L.169S} Strohmayer, T.~E.\ 2001b, \apjl, 554, L169 

\bibitem[Tao et al.(2011)]{2011ApJ...737...81T} Tao, L., Feng, H., Gris{\'e}, F., \& Kaaret, P.\ 2011, \apj, 737, 81 

\bibitem[Timmer \& Koenig(1995)]{1995A&A...300..707T} Timmer, J., \& Koenig, M.\ 1995, \aap, 300, 707 

\bibitem[Tombesi et al.(2010)]{2010A&A...521A..57T} Tombesi, F., Cappi, M., Reeves, J.~N., et al.\ 2010, \aap, 521, A57 

\bibitem[van der Klis(1989)]{1989tns..conf...27V} van der Klis, M.\ 1989, Timing Neutron Stars, 27 

\bibitem[Vaughan \& Uttley(2005)]{2005MNRAS.362..235V} Vaughan, S., \& Uttley, P.\ 2005, \mnras, 362, 235 

\bibitem[Yang et al.(2011)]{2011ApJ...733..118Y} Yang, L., Feng, H., \& Kaaret, P.\ 2011, \apj, 733, 118 

\bibitem[Zhou et al.(2015)]{2015ApJ...798L...5Z} Zhou, X.-L., Yuan, W., Pan, H.-W., \& Liu, Z.\ 2015, \apjl, 798, L5 

\bibitem[Zoghbi et al.(2015)]{2015ApJ...799L..24Z} Zoghbi, A., Miller, J.~M., Walton, D.~J., et al.\ 2015, \apjl, 799, L24 

\end{thebibliography}
\end{document}